\documentstyle[aps]{revtex}
\begin{document}
\author{D. Foerster, CPTMB, Universit\'e de Bordeaux I}
\address{33405 Talence, France}
\author{R. Hayn, Institut f\"{u}r Festk\"{o}rper- und Werkstofforschung (IFW)}
\address{01171 Dresden, Germany}
\author{T. Pruschke and M. Z\"{o}lfl, Institut f\"{u}r Theoretische Physik, Universit\"at Regensburg}
\address{93040 Regensburg, Germany}
\author{H. Rosner, Department of Physics, University of California}
\address{95616 Davis, California, USA}
\title{Metal insulator transition in TlSr$_{2}$CoO$_{5}$ from orbital
degeneracy 
and spin disproportionation}
\maketitle

\begin{abstract}
To describe the metal insulator transition in the new oxide TlSr$_{2}$CoO$%
_{5}$ we investigate the electronic structure of its high temperature
tetragonal phase by local density band structure (LDA) and model
Hartree-Fock calculations. Within LDA we find a homogeneous metallic and
ferromagnetic ground state, but when including the strong Coulomb
interaction in the 3$d$ shell more explicitly within the Hartree-Fock
approximation, we find an insulating state of lower energy that exhibits
both spin and orbital order. The instability of the metallic state towards
the insulating one is driven by orbital degeneracy and a near degeneracy in
energy of the states of intermediate ($s$=1) and high ($s$=2) spin. We also
interpret our results in terms of a simple model. 
\end{abstract}

\subsection*{Introduction}

Six years ago, the observation of colossal magneto resistance in doped
manganites has lead to renewed interest in transition metal oxides, see \cite
{Tokura} for a review. These materials often exhibit complex phases with
coexisting magnetic and orbital \cite{Khomskii} order and with Jahn-Teller
like distortions. The origin of this complexity is the subtle interplay
between different types of local Coulomb interactions within their $3d$
electronic orbitals.

The Co$^{3+}$ analogue TlSr$_{2}$CoO$_{5}$ of non superconducting TlSr$_{2}$%
CuO$_{5}$ is a recent addition to this class of materials. It is of
perovskite structure with a first order transition at $\approx 310$ K from a
tetragonal and metallic phase with ferromagnetic Curie-like susceptibility
at high temperature to an orthorhombic and insulating phase with two
inequivalent kinds of Co ions at low temperature \cite{Coutanceau}. Below $%
T\approx 150$ K antiferromagnetic long range order sets in. Special interest
in this new oxide is due to its quasi two dimensional (2D) character and,
from the point of view of applications, it may be useful that the metal
insulator transition in it occurs at room temperature.

In this paper we explain the 310 K metal insulator transition of TlSr$_{2}$%
CoO$_{5}$ in terms of an instability of its high temperature electronic
structure. We show that the homogeneous ground state can lower its energy by
breaking orbital degeneracy and by acquiring simultaneously both orbital and
spin order.

\subsection*{Method of analysis}

\subsubsection*{(i) Band structure calculation}

We start with a density functional band structure calculation in the
high-temperature phase using the local density approximation (LDA) for the
exchange and correlation potential, see \cite{Kohn} for a review. Above the
transition temperature of 310 K, the crystal structure is tetragonal with
P4/mmm space group (see Fig.\ 1). As input parameters of the calculation, we
used experimentally determined lattice parameters of $a=0.375$ nm, $c=0.877$
nm and relative strontium and apex oxygen (01) positions (that are not fixed
by the space group) of $z=0.2903$ and $z=0.2330$, respectively \cite
{Coutanceau}. We employed a recently developed full-potential nonorthogonal
local-orbital (FPLO) minimum basis scheme \cite{Koepernik} that imposes no
shape restriction on the potential. The calculation was scalar-relativistic,
with the spatial extent of the basis orbitals controlled by a confining
potential of $(r/r_{0})^{4}$ that was optimized with respect to the total
energy. The thallium \{5$s$, 5$p$, 5$d$, 6$s$, 6$p$\}, strontium \{4$s$, 4$p$%
, 5$s$, 5$p$, 5$d$\}, cobalt \{3$s$, 3$p$, 4$s$, 4$p$, 3$d$\} and oxygen \{2$%
s$, 2$p$, 3$d$\} orbitals were treated as valence orbitals, while the lower
lying orbitals were treated as core states. The inclusion of Tl \{5$s$, 5$p$%
, 5$d$\}, Sr \{4$s$, 4$p$\} and Co \{3$s$, 3$p$\} states in the valence
states was necessary to account for non-negligible core-core overlaps. The O
3$d$ states were taken into account to increase the completeness of the
basis set. The results of this LDA calculation will be discussed in detail
further below.

\subsubsection*{(ii) Tight-binding Hartree-Fock calculation}

To take into account the strong Coulomb interactions in the Co $d$ shell
more accurately we could have used the LDA + U approach described in detail
in \cite{Anissimov}. Instead, and in order to gain a more direct insight
into the mechanism of the metal insulator transition at work, we included
the Coulomb interactions in a minimal tight binding Hamiltonian of all
cobalt 3$d$ and oxygen 2$p$ orbitals in the CoO$_{2}$ plane and used the
results of the previously described LDA calculation to determine the hopping
parameters of this model (see \cite{Mizokawa} for similar work). According
to Koster-Slater type symmetry considerations \cite{KosterSlater}, the
overlap between $d$- and $p$-orbitals may be parametrized in terms of two
parameters which we denote as $t_{e_{g}}$ and $t_{t_{2g}}$, respectively.
Due to the perovskite structure, there is no direct Co-Co hopping and we
assume absence of direct O-O hopping. With these approximations, the
transfer matrix elements for the planar Co-O bond in $x$-direction are as
follows 
\begin{equation}
\begin{tabular}{r|@{\ }ccc}
&  & $p^{\vec{x}}$ & $d$ \\ \hline
&  &  &  \\ 
&  &  &  \\ 
$d$ &  & $t_{dp}(\vec{x})$ &  \\ 
&  &  &  \\ 
&  &  & 
\end{tabular}
\quad =\quad 
\begin{tabular}{r|@{\ }ccc}
& $p_{x}^{\vec{x}}$ & $p_{y}^{\vec{x}}$ & $p_{z}^{\vec{x}}$ \\ \hline
$d_{xy}$ & $0$ & $t_{t_{2g}}$ & $0$ \\ 
$d_{yz}$ & $0$ & $0$ & $0$ \\ 
$d_{zx}$ & $0$ & $0$ & $t_{t_{2g}}$ \\ 
$d_{z^{2}}$ & $-\frac{1}{2}t_{e_{g}}$ & $0$ & $0$ \\ 
$d_{x^{2}-y^{2}}$ & $\frac{\sqrt{3}}{2}t_{e_{g}}$ & $0$ & $0$%
\end{tabular}
\label{e1}
\end{equation}
The superscript $\vec{x}$ for the $p$-orbitals indicates that they belong to
the oxygen at $\frac{1}{2}a\vec{e}_{x}$ and the corresponding matrix for the
Co-O bond in $y$-direction is obtained by formally interchanging $%
x\leftrightarrow y$ in the above table. The results of the band structure
calculation determine also the Co $d$ shell crystal-field parameters $%
\epsilon _{i}$, $i=\{xy,yz,zx,z^{2},x^{2}-y^{2}\}$ and the on site energy $%
\epsilon _{p}$ of the oxygen $p$ orbitals (we neglect the splitting of the
oxygen $p$-states due to tetragonal distortion). Tetragonal symmetry at the
Co site leaves $\epsilon _{yz}=\epsilon _{zx}$ degenerate. The sum of
on-site energies and of hopping amplitudes in the $x$- and $y$-directions
defines the noninteracting part of a tight-binding model 
\begin{eqnarray}
H &=&H_{tb}+H_{Coulomb}  \nonumber   \\ \label{e2}
H_{tb} &=&\sum_{i,j,\vec{R},\vec{R}^{\prime },s}t_{ij}(\vec{R}-\vec{R}%
^{\prime })\;a_{i,s}^{\dagger }(\vec{R})\;a_{j,s}(\vec{R}^{\prime })
\end{eqnarray}
where the orbital indices $i,j$ range from 1 to 11 ($1\ldots 5$ are the Co 3$%
d$ orbitals) and $s$ denotes the spin component. To complete the model, we
add the local Coulomb energy at the Co sites as expressed in tight-binding
orbitals 
\begin{equation}
H_{Coulomb}=\frac{1}{2}\sum_{i,j,k,l,s,s^{\prime }}V_{ijkl}a_{i,s}^{\dagger
}a_{j,s^{\prime }}^{\dagger }a_{l,s^{\prime }}a_{k,s}\mbox{ with }%
V_{ijkl}=<ij|\frac{e^{2}}{|\vec{r}-\vec{r}^{\;\prime }|}|kl>
\end{equation}
To express the $V_{ijkl}$ coefficients we use the rotational symmetry of the
Coulomb matrix $\tilde{V}_{ijkl}$ in terms of spherical harmonics $Y_{km}$ 
\cite{Sugano} 
\begin{eqnarray}
\tilde{V}_{m_{1}m_{2}n_{1}n_{2}} &=&\sum_{k=0}^{\infty }\frac{4\pi }{2k+1}%
\int dr_{1}dr_{2}\frac{r_{<}^{k}}{r_{>}^{k+1}}%
R_{d}^{2}(r_{1})R_{d}^{2}(r_{2})\int d\Omega _{1}Y_{2m_{1}}^{\ast
}Y_{2n_{1}}Y_{km}\int d\Omega _{2}Y_{2m_{2}}^{\ast }Y_{2n_{2}}Y_{km}^{\ast }
\label{e4} \\
&=&25\sum_{k=0,2,4}(-)^{m_{1}+n_{2}}F^{(k)}\left( 
\begin{array}{ccc}
2 & 2 & k \\ 
0 & 0 & 0
\end{array}
\right) ^{2}\left( 
\begin{array}{ccc}
2 & 2 & k \\ 
-m_{1} & n_{1} & m
\end{array}
\right) \left( 
\begin{array}{ccc}
2 & 2 & k \\ 
-n_{2} & m_{2} & m
\end{array}
\right)  \nonumber \\
F^{(0)} &=&A+\frac{7}{5}C\mbox{, }\quad F^{(2)}=49B+7C\mbox{, }\quad F^{(4)}=%
\frac{441}{35}C  \nonumber
\end{eqnarray}
and transform $\tilde{V}$ to $V$ in standard cubic orbitals given by 
\begin{equation}
\left( 
\begin{array}{c}
d_{xy} \\ 
d_{yz} \\ 
d_{zx} \\ 
d_{z^{2}} \\ 
d_{x^{2}-y^{2}}
\end{array}
\right) =\left( 
\begin{array}{ccccc}
\frac{i}{\sqrt{2}} & 0 & 0 & 0 & -\frac{i}{\sqrt{2}} \\ 
0 & \frac{i}{\sqrt{2}} & 0 & \frac{i}{\sqrt{2}} & 0 \\ 
0 & -\frac{1}{\sqrt{2}} & 0 & \frac{1}{\sqrt{2}} & 0 \\ 
0 & 0 & 1 & 0 & 0 \\ 
-\frac{1}{\sqrt{2}} & 0 & 0 & 0 & -\frac{1}{\sqrt{2}}
\end{array}
\right) \left( 
\begin{array}{c}
Y_{2-2} \\ 
Y_{2-1} \\ 
Y_{20} \\ 
Y_{21} \\ 
Y_{22}
\end{array}
\right)
\end{equation}
In (\ref{e4}), $R_{d}(r)$ is the radial part of the 3$d$ wave function, $%
\{A,B,C\}$ are Racah parameters and we used formula C.16, p.\ 1057 of \cite
{Messiah}. Please note that expression (\ref{e4}) is rotationally invariant
and that all Coulomb interactions were included. Although we used the full
expression for $V_{ijkl}$ in our calculation, the Racah parameter $B$ is
rather small and when it is taken to be zero we find a much simpler
expression (in the cubic basis) 
\begin{equation}
\left( V_{ijkl}\right) _{B=0}=C\left( \delta _{ij}\delta _{kl}+\delta
_{il}\delta _{jk}\right) +\left( A+C\right) \delta _{ik}\delta _{jl}
\end{equation}
Usually, one denotes the diagonal elements as $U=V_{iiii}=A+3C$ and the
Coulomb repulsion between different ($i\neq j$) orbitals $V_{ijij}=A+C$ is
smaller. However, in our context, it is more convenient to express the
Coulomb energy in terms of the total electronic charge $N=\sum_{i}n_{i}$ and
total electronic spin $\vec{s}_{tot}=\sum_{i}\vec{s}_{i}$ per site: 
\begin{eqnarray}
\left( H_{Coulomb}\right) _{B=0} &\stackrel{\cdot }{=}&C\sum_{ij}P_{i}^{%
\dagger }P_{j}-C\sum_{ij}\vec{s}_{i}\cdot \vec{s}_{j}+\frac{2A+C}{4}%
\sum_{ij}n_{i}n_{j}  \nonumber   \\ \label{e5}
&\approx &J_{H}\left[ D-\left( \vec{s}_{tot}\right) ^{2}\right] +u\frac{N^{2}%
}{2} \\
\mbox{with }\quad n_{i} &=&a_{i,\uparrow }^{\dagger }a_{i,\uparrow
}+a_{i,\downarrow }^{\dagger }a_{i,\downarrow }\;,\quad P_{i}=a_{i,\uparrow
}a_{i,\downarrow }\quad \mbox{and}\quad D=\sum_{i}P_{i}^{\dagger }P_{i}\;, 
\nonumber
\end{eqnarray}
with $u=A+C/2$, $J_{H}=C$ and where $\stackrel{\cdot }{=}$ indicates that
terms that merely renormalize the chemical potential were ignored. In the
last line we retained only the diagonal contribution of the pair hopping
terms.

Treating the Coulomb interaction in an unrestricted HF approximation \cite
{Mermin} that allows for all spin and charge conserving correlations gives
the following renormalization of the on-site Hamilton matrix 
\[
\Delta t_{ij}^{s}=\sum_{l,m}V_{iljm}\sum_{s^{\prime }}n_{lm}^{s^{\prime
}}-\sum_{l,m}V_{ilmj}n_{lm}^{s} 
\]
where $i$,$j$ denote the orbital indices of Co 3$d$ and the density matrix 
\[
n_{lm}^{s}=\frac{1}{N_{k}}\sum_{k,\alpha }^{occ}\psi _{l}^{\ast \alpha s}(%
\vec{k})\psi _{m}^{\alpha s}(\vec{k}) 
\]
is calculated from the normalized wave functions $\psi _{m}^{\alpha s}(\vec{k%
})$ with band index $\alpha $ and momentum $\vec{k}$ ($N_{k}$ corresponds to
the number of $k$ points). The total energy is given by 
\[
E=<H_{tb}>+<H_{Coulomb}> 
\]
where 
\[
<H_{tb}>=\frac{1}{N_{k}}\sum_{k,\alpha ,s}^{occ}\sum_{l,m}\psi _{l}^{\ast
\alpha s}(\vec{k})t_{lm}(\vec{k})\psi _{m}^{\alpha s}(\vec{k}) 
\]
is the mean (unrenormalized) kinetic energy, and 
\begin{equation}
<H_{Coulomb}>=\frac{1}{2}\sum_{m_{1},m_{2},m_{3},m_{4},s,s^{\prime
}}V_{m_{1}m_{2}m_{3}m_{4}}n_{m_{4}m_{2}}^{s}n_{m_{3}m_{1}}^{s^{\prime }}-%
\frac{1}{2}%
\sum_{m_{1},m_{2},m_{3},m_{4},s}V_{m_{1}m_{2}m_{3}m_{4}}n_{m_{4}m_{1}}^{s}n_{m_{3}m_{2}}^{s}
\end{equation}
is the interaction energy.

\subsection*{Results}

\subsubsection*{(i) Full-potential local-orbital method}

We performed two band structure calculations, one spin symmetric (see Fig.\
2) and the other one allowing for spin polarization (Fig.\ 3). Both
solutions are metallic but the magnetic one is energetically preferred by
0.54 eV per formula unit. The density of states (DOS) shows a high degree of
covalency such that part of the magnetic moment (calculated with all the
overlap contributions) sits on oxygen ($m_{O}=0.2\mu _{B}$) giving a total
moment $m=2.1\mu _{B}$ ($m_{Co}=1.9\mu _{B}$). The corresponding occupation
numbers are $n_{Co}=7.2$ and $n_{O}=5.1$.

The band structure of the nonmagnetic solution together with the DOS for an
easier identification of the structures is shown
in Fig.\ 4; the size of the dots included in the band structure in Fig.\ 4 
symbolizes the relative cobalt $3d$-weight in the band.
Evidently, the five bands crossing the Fermi level have predominantly Co 3$d$
character. They
hybridize quite strongly with 2$p$ bands of the in-plane O(2) located at
about 4 to 6 eV binding energy. The 2$p$ bands of the other oxygens O(1) and
O(3) are nonbonding and located in between Co-3$d$ and O(2)-2$p$. The Co-3$d$
bands have only a small dispersion in $z$-direction (with the exception of
Co 3$d_{z^{2}}$) confirming the 2D character of the compound under
consideration. To determine the energetic order of the 3$d$ orbitals we used
the bands at the $\Gamma $ point plus their predominant orbital character 
\cite{Rosner98} and found, in increasing order: $d_{xy}$, $d_{zx/yz}$, $%
d_{z^{2}}$, $d_{x^{2}-y^{2}}$. Due to tetragonal symmetry, there is exact
degeneracy between the $d_{zx}$ and the $d_{yz}$ orbital in the LDA
calculation.

The density of states of the ferromagnetic solution (Fig.\ 3) indicates a
finite value for the majority spin part. From this we conclude that TlSr$%
_{2} $CoO$_{5}$ is not a half-metal and that it may conduct electric current
even in the spin disordered phase. An orbital analysis (that is not
documented here) shows that for majority spin only 3$d_{x^{2}-y^{2}}$ is
partly occupied, while all the remaining majority bands (3$d_{xy}$, 3$%
d_{zx/yz}$, 3$d_{z^{2}}$) are completely below the Fermi level. For minority
spin, 3$d_{xy} $, 3$d_{zx/yz}$ and 3$d_{z^{2}}$ are partly filled, whereas 3$%
d_{x^{2}-y^{2}} $ is nearly empty. It is difficult to interpret such an
itinerant ferromagnet in an ionic picture. But its magnetic moment $m=2s\mu
_{B}>2\mu _{B}$ {\bf \ }indicates a state between intermediate ($s=1$) and
high spin ($s=2$).

\subsubsection*{(ii) Tight-binding Hartree-Fock calculation}

We used the nonmagnetic LDA solution of Fig.\ 4 to fix the parameters of the
Hartree-Fock model (the magnetic solution contains the same information,
except that the spin up and spin down bands are shifted against each other).
By comparing the five relevant bands of the FPLO result (Fig.\ 4) with the
nonmagnetic HF solution we estimated the crystal field parameters to be (in
eV): $\epsilon _{xy}=-1.0$, $\epsilon _{zx/yz}=-0.5$, $\epsilon
_{z^{2}}=-0.2 $, and $\epsilon _{x^{2}-y^{2}}=0.5$. The bandwidth determines
the transfer terms $t_{e_{g}}=1.9$ and $t_{t_{2g}}=1.4$. The Racah
parameters $B$ and $C$ are fixed to their ionic values for trivalent Co as
determined by infrared spectroscopy ($B=0.06$, $C=0.46$) \cite{Ballhausen},
whereas $A=2$ was \ chosen as a typical value for Co. The choice of $%
\epsilon _{p}$ depends on $A $ due to the mean field shift of the 3$d$ level
and we used $\epsilon _{p}=11 $ such that the position of the oxygen levels
(visible as the lower part in Fig.\ 6) coincides with the O(2)-2$p$ position
in LDA. In fact, the position of the oxygen bands does not change very much
for different Hartree-Fock solutions. The tight-binding bands resulting from
these 
parameters are shown in Fig.\ 5 and agree reasonably well with the
nonmagnetic LDA band structure of Fig.\ 4.

Concerning the values of the the crystal field parameters, we note that due
to tetragonal symmetry, there is exact degeneracy between the $d_{zx}$ and $%
d_{yz}$ orbitals in the LDA calculation. However, the values $\varepsilon
_{xy}=-1.0$, $\varepsilon _{zx/yz}=-0.5$ seem to violate ''standard lore'' 
\cite{Sugano} according to which the elongation of the octahedra in the $z$
direction should shift $d_{xy}$ to higher energies compared to $d_{zx/yz}$.
One should note, however, that the $\varepsilon _{i}$ are approximately the
energies at the $\Gamma $ point whereas the ''standard lore'' is valid in
the ionic picture and would correspond, in our case, to the center of
gravity of the different bands. Due to the larger bandwidth of the ''two
dimensional'' $d_{xy}$ band in comparison with the ''one dimensional'' $%
d_{zx/yz}$ bands the corresponding centers of gravity nearly coincide in
Figs.\ 4 and 5. For the electronic structure of the metallic, ferromagnetic
high 
temperature phase of TlSr$_{2}$CoO$_{5}$ it is crucial that all three bands
\{$d_{xy}$, $d_{zx/yz}$\} cross the Fermi level which is due to their widths
being larger than their crystal field splittings, see Fig.\ 4. It
remains an open question, however, how the difference in dimensionality and
width among the \{$d_{xy}$, $d_{zx/yz}$\} bands affects the screening of the
Coulomb interaction. This question was considered in detail \cite
{Lichtenstein} in the context of Sr$_{2}$RuO$_{4}$, another layered
perovskite.

We now discuss the magnetic solutions of the HF approach. The homogeneous
ferromagnetic one (see Table I) is metallic and 650 meV lower in energy than
the nonmagnetic solution. The good agreement with the energy gain in LDA
(540 meV) and similar occupation numbers in LDA and HF indicate that our
parameter assignment is satisfactory. Allowing for different occupations of $%
d_{zx}$ and $d_{yz}$ in a chess board like pattern, we find a metastable
state with orbital order that decays into a ground state with both orbital
and spin order and which contains Co sites with two different
configurations, $m_{B}=3.11\mu _{B}$ (close to high spin) and $m_{A}=2.07\mu
_{B}$ (intermediate spin). The energy gain due to the combined orbital and
spin order is 67 meV. The origin of the instability of the homogeneous
ferromagnetic state is the orbital degeneracy of $d_{zx}$ and $d_{yz}$ and
the near degeneracy of intermediate and high spin configurations. This
instability occurs in a rather large parameter region near the values which
were derived for TlSr$_{2}$CoO$_{5}$. However, due to the large
dimensionality of the parameter space of our HF model (\ref{e2}), we did not
perform a systematic study. In Fig.\ 6 we give the spectral density of the
ferromagnetic and of the two chess board like ordered states. As we can see,
any kind of order leads to a decrease of spectral density at the Fermi
level, but only the orbital- and spin-ordered solution is insulating. That
is also visible in the HF band structure (Fig.\ 7).

The chess board like superstructure found here does not correspond to the
orthorhombic low-temperature phase seen experimentally with a 2:1 ratio of
high spin and intermediate spin states \cite{Coutanceau}. We also
investigated this experimental superstructure of the CoO$_{2}$ plane within
the model Hartree-Fock approach but found no solution with lower energy than
those with chess board order. The lattice degrees of freedom may have to be
included into the model to obtain the correct pattern, because the nearest
neighbor Co-O distance for high spin is probably larger than the
corresponding distance for intermediate spin (as can be concluded from the
analogy to the famous Invar alloys Fe$_{x}$Ni$_{1-x}$ \cite{Wassermann}).
The present investigation of the electron system alone can only indicate the
instability of the high temperature phase but it is not able to predict the
correct low-temperature crystal structure.

\subsection*{Interpretation of results}

We now interpret the results of our LDA and HF calculations in terms of a
simplified model. As a first step, we recall that there is only $%
d\leftrightarrow p$ hopping and neither direct $d\leftrightarrow d$ nor $%
p\leftrightarrow p$ hopping. This and the fact that only cobalt $d$ states
are at the Fermi level allows us to extract an effective $d\leftrightarrow d$
hopping by eliminating the oxygen orbitals in standard fashion \cite{Fulde} 
\begin{eqnarray}  \label{e7}
t_{dd^{\prime }}^{eff}(\vec{a}) &=& \varepsilon_{d}\delta _{dd^{\prime }} +%
\frac{t_{dp} (\vec{a}) t_{d^{\prime }p} (-\vec{a})}{\Delta \varepsilon}
=\varepsilon_{d}\delta_{dd^{\prime }} -\frac{t_{dp} (\vec{a})t_{pd^{\prime
}} (\vec{a})}{\Delta \varepsilon} \\
t_{dd^{\prime }}^{eff} (\vec{x}) &=&\varepsilon _{d}\delta _{dd^{\prime }}- 
\frac{1}{\Delta \varepsilon} \; \left( 
\begin{array}{ccccc}
t_{t_{2g}}{}^{2} & 0 & 0 & 0 & 0 \\ 
0 & 0 & 0 & 0 & 0 \\ 
0 & 0 & t_{t_{2g}}{}^{2} & 0 & 0 \\ 
0 & 0 & 0 & \frac{1}{4}t_{e_{g}}{}^{2} & -\frac{\sqrt{3}}{4}t_{e_{g}}{}^{2}
\\ 
0 & 0 & 0 & -\frac{\sqrt{3}}{4}t_{e_{g}}{}^{2} & \frac{3}{4}t_{e_{g}}{}^{2}
\end{array}
\right)  \nonumber
\end{eqnarray}
where $\vec{a}$ $\epsilon \{\vec{x},\vec{y},-\vec{x},-\vec{y}\}$ denotes the
direction of hopping and $\Delta \varepsilon$ is the offset between the Co
and oxygen bands. Actually, due to the high degree of $d\leftrightarrow p$
hybridization the second order expression (\ref{e7}) is certainly not
sufficient to provide correct numbers for $t^{eff}_{dd^{\prime}}$ but
nonetheless it should give the correct matrix structure. Just like the
original hopping the effective $t^{eff}_{dd^{\prime}}$ is also anisotropic
and orbitally dependent, with $d_{zx}$ and $d_{yz}$ electrons hopping in the 
$x$ respectively $y$ direction and forming one dimensional bands. Using the
Coulomb energy of Eq.\ (\ref{e5}) this provides us with a simplified model
that involves only $d$ orbitals: 
\begin{equation}
H=\sum_{d,s,\vec{R}} \varepsilon _{i}a_{d,s}^{\dagger}(\vec{R}) a_{d,s} (%
\vec{R}) +\sum_{d,d^{\prime},\vec{R},\vec{R}^{\prime},s}
t_{dd^{\prime}}^{eff} (\vec{R}-\vec{R}^{\prime}) \; a_{d,s}^{\dagger} (\vec{R%
}) \; a_{d^{\prime},s} (\vec{R}^{\prime}) +\sum_{\vec{R}}j_{H}\left[ D_{\vec{%
R}}-\left( \vec{s}_{tot,\vec{R}}\right)^{2}\right] +\frac{u}{2}N_{\vec{R}%
}^{2}  \label{e8}
\end{equation}
where $j_{H}$ and $u$ are renormalized values of $J_{H}$ and $\tilde U$ due
to the elimination of the oxygen orbitals. But the precise amount of
renormalization is difficult to calculate and beyond the scope of the
present discussion.

The above five band model must be simplified further to extract the relevant
degrees of freedom. We first note that the crystal field parameters derived
in the previous section are such that in the ionic case ($t_{dd^{\prime
}}^{eff}=0$) the three configurations of Fig.\ 8 are lowest in energy. We
further highlight the orbitals closest to the Fermi level and which we
believe to be itinerant by bold arrows. This suggests a minimal model with
three itinerant electron species, namely $d_{x^{2}-y^{2}}^{\uparrow }$ and
the two degenerate minority spin bands $d_{zx}^{\downarrow }$ and $%
d_{yz}^{\downarrow }$. This physical picture is also supported by the
ferromagnetic HF solution where only $d_{x^{2}-y^{2}}^{\uparrow }$ is partly
occupied among all the majority spin bands and the occupation of $%
d_{z^{2}}^{\downarrow }$ and $d_{x^{2}-y^{2}}^{\downarrow }$ in Table I is
indeed small (the occupation of $d_{xy}^{\downarrow }$, however, deviates
quite strongly from unity). The low energy sector responsible of the metal
insulator and spin transition should be the competition between 
\begin{eqnarray}
s &=&1:\quad xy\uparrow \downarrow ,\mbox{ }zx\uparrow \downarrow ,\mbox{ \ }%
yz\uparrow ,\mbox{ }3z^{2}-r^{2}\uparrow \mbox{or }x\leftrightarrow y \\
s &=&2:\quad xy\uparrow \downarrow ,\mbox{ }zx\uparrow ,\mbox{ \ }yz\uparrow
,\mbox{ }3z^{2}-r^{2}\uparrow ,x^{2}-y^{2}\uparrow   \nonumber
\end{eqnarray}
In other words, $xy\uparrow \downarrow $ is only a spectator orbital while \{%
$zx\uparrow ,$ \ $yz\uparrow ,$ $3z^{2}-r^{2}\uparrow \}$ provide a total
spin of 3/2 and the exact degeneracy between $d_{zx}$ and \ $d_{yz}$ is kept
as one of the driving mechanisms of the transition. The spin transition is
then due to the competition between a down spin electron $zx\downarrow $ or $%
yz\downarrow $ and an up spin electron $x^{2}-y^{2}\uparrow $. Combining the
spins of \{$zx\uparrow ,$ \ $yz\uparrow ,$ $3z^{2}-r^{2}\uparrow \}$ into an
effective spin 3/2 degree of freedom, we can represent this competition as
follows: 
\begin{equation}
\{S_{\frac{3}{2}}^{\uparrow }\mbox{, }zx\downarrow \}\quad \mbox{or}\quad
\{S_{\frac{3}{2}}^{\uparrow }\mbox{, }yz\downarrow \}\quad \Leftrightarrow %
\mbox{ \ \ }\{S_{\frac{3}{2}}^{\uparrow }\mbox{, }x^{2}-y^{2}\uparrow \}
\end{equation}
where we used Hund's coupling to exclude misaligned spins. To describe these
qualitative ideas more precisely, we propose the following model: 
\begin{eqnarray}
H &=&\varepsilon \sum_{\vec{R},s}a_{3,s}^{\dagger }(\vec{R})\;a_{3,s}(\vec{R}%
)-\sum_{\vec{R},\vec{R}^{\prime },s}\sum_{i=1}^{3}t_{ii}^{eff}(\vec{R}-\vec{R%
}^{\prime })\;a_{i,s}^{\dagger }(\vec{R})\;a_{i,s}(\vec{R}^{\prime })
\label{e11} \\
&&-2j_{H}\sum_{\vec{R}}\vec{S}_{\vec{R}}\left( \vec{s}_{\vec{R},3}-\vec{s}_{%
\vec{R},2}-\vec{s}_{\vec{R},1}\right) +\frac{u}{2}\sum_{\vec{R},i,j}n_{\vec{R%
},i}n_{\vec{R},j}+j\sum_{\vec{R},\vec{R}^{\prime }}\vec{S}_{\vec{R}}\cdot 
\vec{S}_{\vec{R}^{\prime }}  \nonumber
\end{eqnarray}
where the orbitals are numbered as $\{1,2,3\}\leftrightarrow
\{zx,yz,x^{2}-y^{2}\}$ and $t_{ii}^{eff}$ is an anisotropic and orbitally
dependent hopping matrix. Model (\ref{e11}) is formulated in an extended
phase space in comparison to (\ref{e8}) since $|S_{\frac{3}{2}}^{\uparrow
},zx\uparrow \rangle $ and $|S_{\frac{3}{2}}^{\uparrow },yz\uparrow \rangle $
do not exist in the 5 band model. But those unphysical states are at a high
energy due to Hund's coupling $j_{H}$. The origin of the antiferromagnetic
exchange coupling $j$ should be the virtual superexchange of the singly
occupied states that were excluded from our model and the coupling $\vec{S}_{%
\vec{R}}\left( \vec{s}_{\vec{R},3}-\vec{s}_{\vec{R},2}-\vec{s}_{\vec{R}%
,1}\right) $ makes sure that the spins of electrons in the orbitals $i=1,2$
are antiparallel to the spin 3/2 vector and vice versa for $i=3$. A detailed
estimate of the model parameters in (\ref{e11}) is beyond the scope of the
present paper, but it is clear that we have to consider the range of
parameters $u\gg j_{H}\gg \varepsilon ,t,j$. The condition $j_{H}\gg t$
suppresses virtual hopping processes for antiparallel nearest neighbor spins 
$\vec{S}_{\vec{R}}$, whereas they remain possible for parallel spins. In the
limit of $j_{H}\gg \varepsilon ,t$ the fermions are spin polarized, double
occupation of the same orbital is automatically excluded by Fermi statistics
and we therefore dropped the local double occupation $D$-term in equation 
(\ref{e11}). The above model still admits high energy processes (involving
energy costs of $u$ and $j_{H}$) that must be integrated out to obtain a
true low energy model.

To argue for the minimal model (\ref{e11}) we compare some of the possible
two site clusters in second order perturbation theory (Fig.\ 9) (for
simplicity we use identical transfer amplitudes $t$ and restrict ourselves
to leading terms). We see that for a ferromagnetic spin arrangement there is
a competition between an orbitally ordered state $s=1$ (realized for $%
\varepsilon > t^2/u$) and a mixed orbital and spin ordered state ($%
\varepsilon < t^2/u$) which is also realized in the model HF approach. Next,
we see from the mixed spin cluster in the figure that hopping processes are
influenced by the relative spin orientation. In this cluster, there is a
competition between ferromagnetic order favored by a gain in delocalization
energy $\sim \frac{t^2}{u}\frac{j_H}{u}$ and antiferromagnetic order favored
by a gain in magnetic energy $\sim j$. The ferromagnetic delocalization
energy is expected to dominate $j$ and therefore neighboring high spin and
intermediate spin states should order ferromagnetically (see \cite
{Coutanceau}) and it is easy to see that neighboring high spin states ($s=2$%
) have an antiferromagnetic exchange in our simplified model (\ref{e11}).

There is a similarity between the model we propose here and the Zener double
exchange model, \cite{Zener}, except that our model has flipped spins for
some of the orbitals to describe the spin transition and we also have a
total of 3 species of electrons per site at $j_{H}\gg t$. For sufficiently
small values of $u$ there will be, by analogy with the double exchange
model, a ferromagnetic and metallic phase because charge transport is
possible for parallel spin orientations, while for $u\to\infty$ the system
is insulating. So we expect a rich phase diagram of this reduced model as a
function of its parameters $\left\{ \varepsilon ,t,j_{H},u,j\right\} $ with
phases of mixed magnetic and orbital or spin order and which may be metallic
or insulating as a function of its parameters.

\subsection*{Conclusions}

Using LDA band structure calculations for the high temperature tetragonal
phase and a HF approach in a minimal model of the perovskite plane, we found
states that are lower in energy than the homogeneous and ferromagnetic
state. More specifically we found that the homogeneous state is unstable
towards orbital order and spin state disproportionation. We propose this
instability to be the driving mechanism of the metal to insulator
transition. A dynamical spin disproportionation above the transition
temperature would be compatible with M\"{o}ssbauer data on $^{57}$Fe-doped
TlSr$_{2}$CoO$_{5}$ that suggest the existence of two inequivalent magnetic
sites \cite{Moessbauer}. Based on our calculations, we also proposed a
simplified model with only three states per site at $j_{H}\gg t$ plus an
extra spin of $S=\frac{3}{2}$ and which we argue to have a rich phase
diagram as a function of its model parameters.

\noindent {\bf Acknowledgements}

\noindent We are indebted to J.-P. Doumerc for many helpful and inspiring
discussions on his data, to A. Villesuzanne for a comparison with his own
band structure results and to D. Khomskii and M. Pouchard for useful
criticism. H. R. acknowledges funding by an individual grant of ''Deutscher
Akademischer Austauschdienst'', T. P. and R.H. both acknowledge support as
''Professeur Invit\'{e} '' and D.F. was supported by ''Graduiertenkolleg
Komplexit\"{a}t in Festk\"{o}rpern'', IFW Dresden and ''Groupement de
Recherche Oxydes Remarquables''. CPTMB is "Equipe Associ\'ee au CNRS ERS 2120".  

After this paper was completed we were kindly informed by C. Michel of an
oxide synthesized at Caen \cite{Raveau} that contains CoO$_{2}$ planes
isostructural to the ones considered here.

\newpage

{\bf Figure Captions}

\vspace*{0.5cm}

\noindent Fig.\ 1: Crystal structure of the high temperature tetragonal
phase. 

\vspace*{0.5cm}

\noindent Fig.\ 2: DOS of the nonmagnetic solution.

\vspace*{0.5cm}

\noindent Fig.\ 3: DOS of the ferromagnetic state.

\vspace*{0.5cm}

\noindent Fig.\ 4: Nonmagneticc LDA band structure. The relative cobalt $3d$
weight of the bands is symbolized via black dots in the bandstructure. For
comparison we also included the DOS (see Fig.\ 2) at the right hand side of
the band structure. The broad line at the Fermi level between $X$ and $M$ is
comprised of 3 bands ($d_{xy}$, $d_{zx}$, and $d_{z^2}$) which are nearly
degenerate. 

\vspace*{0.5cm}

\noindent Fig.\ 5: Band structure of the nonmagnetic HF solution.

\vspace*{0.5cm}

\noindent Fig.\ 6: Comparison of spectral weights for three magnetic HF
solutions.

\vspace*{0.5cm}

\noindent Fig.\ 7: HF band structure of the orbital- and spin-ordered
solution.

\vspace*{0.5cm}

\noindent Fig.\ 8: The lowest ionic configurations of Co$^{3+}$ in TlSr$_2$%
CoO$_5$. Configurations a and b correspond to intermediate spin states and
are degenerate ($d_{zx}^{\downarrow}$,$d_{yz}^{\downarrow}$) whereas c is a
high spin state ($d_{x^2-y^2}^{\uparrow}$). The bold arrows correspond to
itinerant states, the remaining states are localized and are combined into
an effectice spin 3/2 (middle row). The right row introduces a notation.

\vspace*{0.5cm}

\noindent Fig.\ 9: Perturbation theory for configurations of a two site
cluster.

\vspace*{2cm}

\noindent {\bf Table I}: Occupation numbers of HF solutions.

\vspace*{0.5cm}

\begin{center}
\begin{tabular}{c|c|c|c|c|c|c|c|c|c}
Solution & Energy/meV & \multicolumn{7}{c}{Occupation numbers} & 
Magnetization/$\mu_B$ \\ 
&  &  & $d_{xy}$ & $d_{zx}$ & $d_{yz}$ & $d_{z^2}$ & $d_{x^2-y^2}$ & sum & 
\\ \hline
Nonmagnetic & 0 & $n_d$ & 0.84 & 0.81 & 0.81 & 0.76 & 0.30 & 7.04 &  \\ \hline
Ferromagnetic & -653 & $n_d^{\uparrow}$ & 1.00 & 1.00 & 1.00 & 0.98 & 0.63 & 
4.61 & 2.41 \\ 
&  & $n_d^{\downarrow}$ & 0.55 & 0.70 & 0.70 & 0.08 & 0.17 & 2.20 &  \\ 
\hline
Orbital order & -674 & $n_{dA}^{\uparrow }$ & 1.00 & 1.00 & 1.00 & 0.98 & 
0.67 & 4.65 & 2.52 \\ 
&  & $n_{dA}^{\downarrow }$ & 0.69 & 0.89 & 0.29 & 0.09 & 0.17 & 2.13 &  \\ 
(quasistable) &  & $n_{dB}^{\uparrow }$ & 1.00 & 1.00 & 1.00 & 0.98 & 0.67 & 
4.65 & 2.52 \\ 
&  & $n_{dB}^{\downarrow }$ & 0.69 & 0.29 & 0.89 & 0.09 & 0.17 & 2.13 &  \\ 
\hline
Spin and & -720 & $n_{dA}^{\uparrow }$ & 1.00 & 1.00 & 1.00 & 1.00 & 0.45 & 
4.45 & 2.07 \\ 
&  & $n_{dA}^{\downarrow }$ & 0.90 & 0.90 & 0.31 & 0.09 & 0.18 & 2.38 &  \\ 
orbital order &  & $n_{dB}^{\uparrow }$ & 1.00 & 1.00 & 1.00 & 0.99 & 0.91 & 
4.90 & 3.11 \\ 
&  & $n_{dB}^{\downarrow }$ & 0.37 & 0.28 & 0.88 & 0.09 & 0.17 & 1.79 &  \\ 
\hline
\end{tabular}
\end{center}

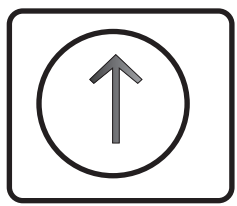
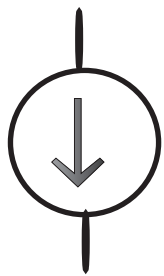
\end{document}